\documentclass[aps,twocolumn,groupedaddress,amsmath,amssymb]{revtex4}

\usepackage{graphicx}% Include figure files
\usepackage{dcolumn}% Align table columns on decimal point
\usepackage{xcolor}%
\usepackage{booktabs}%
\usepackage{bm}% bold math
\usepackage{color, soul}

\begin{document}

\title{High-entropy silicide superconductors with W$_{5}$Si$_{3}$-type structure}

\date{\today}
\author{Bin Liu$^{1}$}
\email{bliu0201@foxmail.com}
\author{Wuzhang Yang$^{2,3,5}$}
\author{Guorui Xiao$^{2,3,4}$}
\author{Qinqing Zhu$^{2,3,5}$}
\author{Shijie Song$^{4}$}
\author{Guang-Han Cao$^{4}$}
\author{Zhi Ren$^{2,3}$}
\email{renzhi@westlake.edu.cn}

\affiliation{$^{1}$Faculty of Materials Science and Engineering, Kunming University of Science and Technology, Kunming, Yunnan 650000, PR China}
\affiliation{$^{2}$Department of Physics, School of Science, Westlake University, 18 Shilongshan Road, Hangzhou, 310024, Zhejiang Province, PR China}
\affiliation{$^{3}$Institute of Natural Sciences, Westlake Institute for Advanced Study, 18 Shilongshan Road, Hangzhou, 310024, Zhejiang Province, PR China}
\affiliation{$^{4}$School of Physics, Zhejiang University, Hangzhou 310058, PR China}
\affiliation{$^{5}$Department of Physics, Fudan University, Shanghai 200433, PR China}

\begin{abstract}
We report the synthesis, crystal structure and physical properties of two new high-entropy silicides (HESs), namely (Nb$_{0.1}$Mo$_{0.3}$W$_{0.3}$Re$_{0.2}$Ru$_{0.1}$)$_{5}$Si$_{3}$ and (Nb$_{0.2}$Mo$_{0.3}$W$_{0.3}$Re$_{0.1}$Ru$_{0.1}$)$_{5}$Si$_{3}$. Structural analysis indicates that both HESs consist of a (nearly) single tetragonal W$_{5}$Si$_{3}$-type phase (space group $I$4/$mcm$) with a disordered cation distribution.
Electrical resistivity, magnetic susceptibility and specific heat measurements show that (Nb$_{0.1}$Mo$_{0.3}$W$_{0.3}$Re$_{0.2}$Ru$_{0.1}$)$_{5}$Si$_{3}$ and (Nb$_{0.2}$Mo$_{0.3}$W$_{0.3}$Re$_{0.1}$Ru$_{0.1}$)$_{5}$Si$_{3}$ are weakly coupled bulk superconductors, which represent the first superconducting high-entropy nonoxide ceramics. In particular, these HESs have higher $T_{\rm c}$ values (3.2-3.3 K) compared with those of the binary counterparts, and their $B_{\rm c2}$(0)/$T_{\rm c}$ ratios are the largest among superconductors of the same structural type.
\end{abstract}

\maketitle
\maketitle

\begin{figure*}
	\includegraphics*[width=17.6cm]{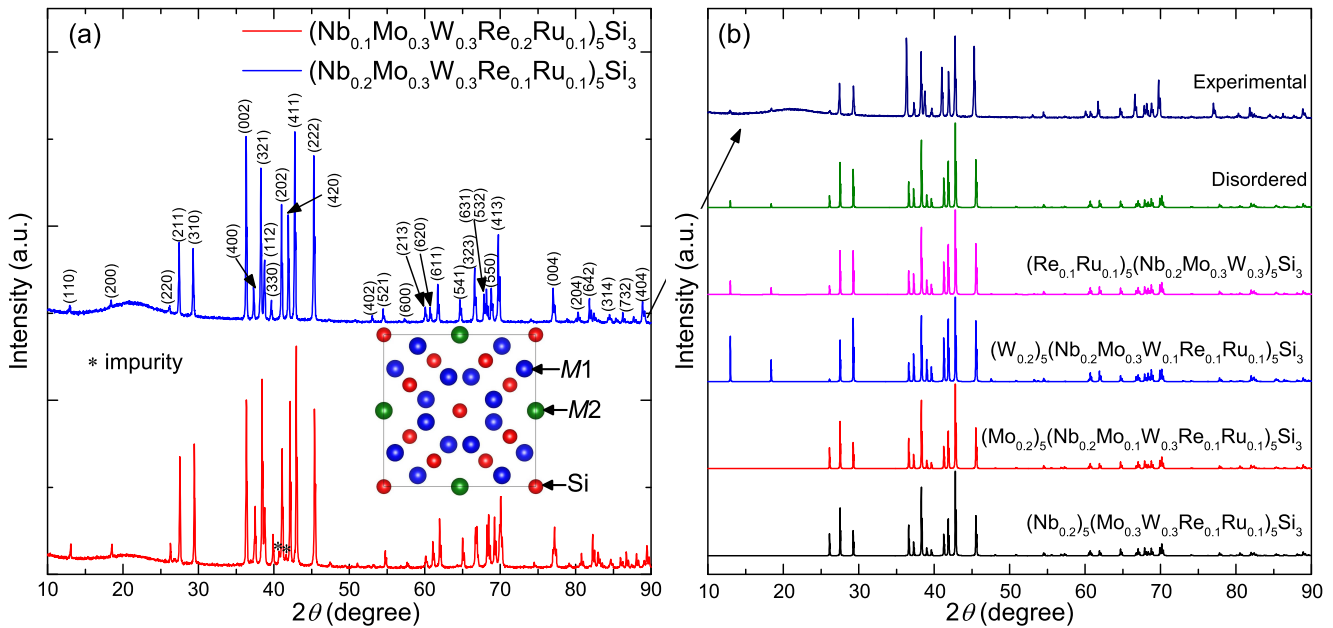}
	\caption{
		(a) XRD patterns for (Nb$_{0.1}$Mo$_{0.3}$W$_{0.3}$Re$_{0.2}$Ru$_{0.1}$)$_{5}$Si$_{3}$ and (Nb$_{0.2}$Mo$_{0.3}$W$_{0.3}$Re$_{0.1}$Ru$_{0.1}$)$_{5}$Si$_{3}$.
         The diffraction peaks for the latter are indexed on a tetragonal lattice with the $I$4/$mcm$ space group.
         The inset shows a schematic structure of M$_{5}$Si$_{3}$ projected perpendicular to the $ab$ plane.
		(b) Comparison between the experimental XRD pattern of (Nb$_{0.2}$Mo$_{0.3}$W$_{0.3}$Re$_{0.1}$Ru$_{0.1}$)$_{5}$Si$_{3}$ and calculated ones based on different structural models.
	}
	\label{fig1}
\end{figure*}

\begin{table*}
	\caption{Structural and physical parameters of (Nb$_{0.1}$Mo$_{0.3}$W$_{0.3}$Re$_{0.2}$Ru$_{0.1}$)$_{5}$Si$_{3}$ and (Nb$_{0.2}$Mo$_{0.3}$W$_{0.3}$Re$_{0.1}$Ru$_{0.1}$)$_{5}$Si$_{3}$.}
	\renewcommand\arraystretch{1.2}
	\begin{tabular}{p{2.6cm}<{\centering}p{6cm}<{\centering}p{6cm}<{\centering}}
		\\
		\hline
		Parameter          & (Nb$_{0.1}$Mo$_{0.3}$W$_{0.3}$Re$_{0.2}$Ru$_{0.1}$)$_{5}$Si$_{3}$ & (Nb$_{0.2}$Mo$_{0.3}$W$_{0.3}$Re$_{0.1}$Ru$_{0.1}$)$_{5}$Si$_{3}$\\
		\hline
		$a$ ({\AA})& 9.644(1)& 9.650(1)\\
        $c$ ({\AA})& 4.948(1)& 4.951(1) \\
        $T_{\rm c}$ (K) & 3.3& 3.2 \\
		$\gamma$ (mJ/molK$^{2}$) & 21.7& 20.8 \\
		$\beta$ (mJ/molK$^{4}$) & 0.2353 &0.1842 \\
		$\Theta_{\mathrm{D}}$ (K)   & 404& 472    \\
		$\lambda_{\mathrm{ep}}$      & 0.53 & 0.51  \\
        $B_{\rm c2}$(0) (T)     & 5.0& 5.1  \\
		$\xi_{\mathrm{GL}}(0)$ (nm)   & 8.1& 8.0  \\
		\hline
	\end{tabular}
	\label{Table1}
\end{table*}

\section{Introduction}
Over the past two decades, the development of functional materials based on the high-entropy concept has attracted a lot of attention \cite{ref1,ref2,ref3,ref4}.
This concept is initially proposed in the field of alloys and later extended to ceramics including oxides, nitrides, carbides, borides, silicides, etc \cite{HEC1,HEC2,HEC3,HEC4,HEC5,HEC6,HEC7,HEC8,HEC9,HEC10}.
By definition, high-entropy ceramics (HECs) are single phase ceramics where the configurational entropy plays a role in their formation and stability but is not necessarily dominant over enthalpy \cite{HEC10}.
So far, the HECs are found to exist in diverse structures, ranging from simple rock-salt type to complex perovskite type, and exhibit a variety of novel physical and mechanical properties, such as
reduced thermal conductivity \cite{ref9}, enhanced thermoelectric performance \cite{ref10}, improved ionic conductivity \cite{ref11} and high hardness \cite{ref12}.
Nonetheless, contrary to high-entropy alloys, there are few HECs that display superconductivity.

Binary transition metal silicides $M_{5}$Si$_{3}$ have been studied extensively for potential high temperature structural applications due to their high melting points \cite{ref13,ref14}.
Depending on the nature of $M$ element, these silicides crystallize in either the tetragonal Cr$_{5}$B$_{3}$-type, W$_{5}$Si$_{3}$-type or hexagonal Mn$_{5}$Si$_{3}$-type structures \cite{ref15}.
Examples falling in the W$_{5}$Si$_{3}$-type category include V$_{5}$Si$_{3}$ \cite{ref16}, $\beta$-Nb$_{5}$Si$_{3}$ \cite{ref17}, Cr$_{5}$Si$_{3}$ \cite{ref18}, Mo$_{5}$Si$_{3}$ \cite{ref19} and W$_{5}$Si$_{3}$ itself \cite{ref20}.
Notably, superconductivity is observed for $\beta$-Nb$_{5}$Si$_{3}$ \cite{ref21} as well as W$_{5}$Si$_{3}$ \cite{ref22}, and can be induced in Mo$_{5}$Si$_{3}$ by partial substitution of Mo with Re \cite{ref23}.
However, no W$_{5}$Si$_{3}$-type high-entropy silicide (HES) has been reported to date.
As a matter of fact, both (V$_{1/5}$Cr$_{1/5}$Nb$_{1/5}$Ta$_{1/5}$W$_{1/5}$)$_{5}$Si$_{3}$ and (Ti$_{1/5}$Zr$_{1/5}$Nb$_{1/5}$Mo$_{1/5}$Hf$_{1/5}$)$_{5}$Si$_{3}$ silicides adopt the hexagonal Mn$_{5}$Si$_{3}$-type structure \cite{ref24}.
This unexpected phase is likely stabilized by the cation ordering, as revealed by structural characterization. Previously, MoSi$_{2}$-type HESs such as (Mo$_{0.2}$Nb$_{0. 2}$Ta$_{0.2}$Ti$_{0.2}$W$_{0. 2}$)Si$_{2}$ and (Mo$_{0.2}$W$_{0.2}$Cr$_{0.2}$Ta$_{0.2}$Nb$_{0.2}$)Si$_{2}$ have also been obtained and studied extensively \cite{MoSi2-1,MoSi2-2,MoSi2-3}.

Here we present the first synthesis and characterization of the (Nb$_{0.1}$Mo$_{0.3}$W$_{0.3}$Re$_{0.2}$Ru$_{0.1}$)$_{5}$Si$_{3}$ and (Nb$_{0.2}$Mo$_{0.3}$W$_{0.3}$Re$_{0.1}$Ru$_{0.1}$)$_{5}$Si$_{3}$ HESs.
The analysis of x-ray diffraction (XRD) patterns indicates that both HESs are of a (nearly) single W$_{5}$Si$_{3}$-type phase, in which the transition metal atoms are distributed disorderly at the cation sites.
Moreover, the (Nb$_{0.1}$Mo$_{0.3}$W$_{0.3}$Re$_{0.2}$Ru$_{0.1}$)$_{5}$Si$_{3}$ and (Nb$_{0.2}$Mo$_{0.3}$W$_{0.3}$Re$_{0.1}$Ru$_{0.1}$)$_{5}$Si$_{3}$ HESs exhibit bulk superconductivity below 3.3 K and 3.2 K, respectively.
Characteristic normal-state and superconducting parameters are obtained and compared with those of isostructural superconductors, the implications of which are discussed.

\section{Materials and Methods}
Polycrystalline (Nb$_{0.1}$Mo$_{0.3}$W$_{0.3}$Re$_{0.2}$Ru$_{0.1}$)$_{5}$Si$_{3}$ and (Nb$_{0.2}$Mo$_{0.3}$W$_{0.3}$Re$_{0.1}$Ru$_{0.1}$)$_{5}$Si$_{3}$ silicides were synthesized by the arc melting method. High purity Nb(99.99\%), Mo(99.99\%), W(99.9\%), Re(99.9\%), Ru(99.9\%) and Si(99.99\%) powders were weighted according to the stoichiometric ratios, mixed thoroughly and pressed into pellets in an argon-filled glove box. Afterwards, the pellets were put into an arc furnace and melted quickly several times under high-purity argon atmosphere, followed by rapid cooling on a water-chilled copper plate. Finally, the as-cast ingots were crushed, pressed again into pellets, which were placed inside the alumina crucibles and sealed in the Ta tubes. The Ta tubes were then annealed at 1600 $^{\circ}$C in a vacuum furnace for 12 h, followed by furnace cooled to room temperature.
The phase purity of the resulting samples was analyzed by powder XRD using a Bruker D8 Advance X-ray diffractometer with Cu K$\alpha$ radiation at room temperature. The morphology and elemental composition were then characterized in a scanning electron microscope (SEM, Supra-55 Zeiss) equipped with energy dispersive spectroscopy (EDS). The microstructure was examined in an FEI Tecnai G2 F20 S-TWIN transmission electron microscope (TEM) operated under an accelerating voltage of 200 kV. The electrical resistivity and specific heat were measured by the standard four-probe and the relaxation methods, respectively, in a Quantum Design Physical Property Measurement System (PPMS-9 Dynacool). The dc magnetization measurements down to 1.8 K were performed in a commercial SQUID magnetometer (MPMS3) with an applied magnetic field of 1 mT in both the zero-field cooling (ZFC) and field cooling (FC) modes.
\begin{figure*}
	\includegraphics*[width=17cm]{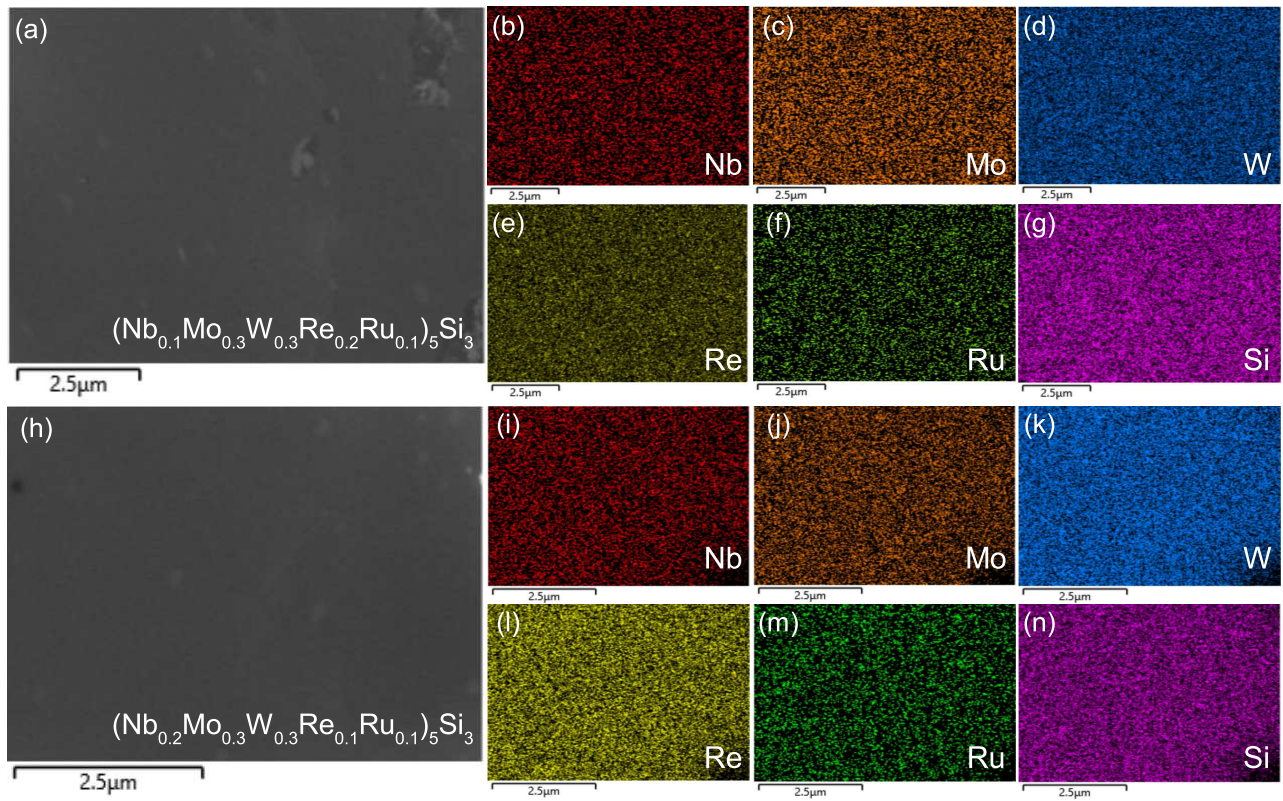}
	\caption{
		(a) SEM image with a scale bar of 2.5 $\mu$m for (Nb$_{0.1}$Mo$_{0.3}$W$_{0.3}$Re$_{0.2}$Ru$_{0.1}$)$_{5}$Si$_{3}$.
        (b-g) EDX mapping results for the Nb, Mo, W, Re, Ru and Si elements, respectively.
        (h-n) Same set of data for (Nb$_{0.2}$Mo$_{0.3}$W$_{0.3}$Re$_{0.1}$Ru$_{0.1}$)$_{5}$Si$_{3}$.
	}
	\label{fig2}
\end{figure*}
\section{Results and Discussion}
Figure 1(a) shows the results of XRD measurements for the annealed
(Nb$_{0.1}$Mo$_{0.3}$W$_{0.3}$Re$_{0.2}$Ru$_{0.1}$)$_{5}$Si$_{3}$ and (Nb$_{0.2}$Mo$_{0.3}$W$_{0.3}$Re$_{0.1}$Ru$_{0.1}$)$_{5}$Si$_{3}$ samples.
The two diffraction patterns are very similar and can be well indexed on the tetragonal W$_{5}$Si$_{3}$-type structure with the $I$4/$mcm$ space group. In addition, two small impurity peaks at 2$\theta$ $\approx$ 40.8$^{\circ}$ and 41.7$^{\circ}$ are identified for (Nb$_{0.1}$Mo$_{0.3}$W$_{0.3}$Re$_{0.2}$Ru$_{0.1}$)$_{5}$Si$_{3}$, which can be assigned to the MoSi$_{2}$-type and bcc-type phases, respectively. It is emphasized that the annealing process is crucial for obtaining the (nearly) single phase since multiple phases are formed in the as-cast state.\cite{SM}
The lattice parameters determined by the least-squares fittings are $a$ = 9.644(1) {\AA}, $c$ = 4.948(1) {\AA} for (Nb$_{0.1}$Mo$_{0.3}$W$_{0.3}$Re$_{0.2}$Ru$_{0.1}$)$_{5}$Si$_{3}$, and $a$ = 9.650(1) {\AA}, $c$ = 4.951(1) {\AA} for (Nb$_{0.2}$Mo$_{0.3}$W$_{0.3}$Re$_{0.1}$Ru$_{0.1}$)$_{5}$Si$_{3}$.
It is noted that the $a$-axis values are close to that of Mo$_{5}$Si$_{3}$ ($a$ = 9.648 {\AA}, $c$ = 4.903 {\AA}) \cite{ref19}, while the $c$-axis ones are comparable to that of W$_{5}$Si$_{3}$ ($a$ = 9.612 {\AA}, $c$ = 4.961 {\AA}) \cite{ref20}.
Nevertheless, both $a$ and $c$-axis of the HESs are shorter than those of $\beta$-Nb$_{5}$Si$_{3}$ ($a$ = 10.02 {\AA}, $c$ = 5.072 {\AA}) \cite{ref17}.
Clearly, the incorporation of Re and Ru tends to shrink the lattice, even though W$_{5}$Si$_{3}$-type Re$_{5}$Si$_{3}$ and Ru$_{5}$Si$_{3}$ do not exist \cite{ref25,ref26}.
This is consistent with their smaller atomic radii compared with the group VB and VIB elements and is reminiscent of the cases in Mo$_{5-x}$Re$_{x}$Si$_{3}$ and W$_{5-x}$Re$_{x}$Si$_{3}$ \cite{ref23,ref27}.
\begin{figure*}
	\includegraphics*[width=17cm]{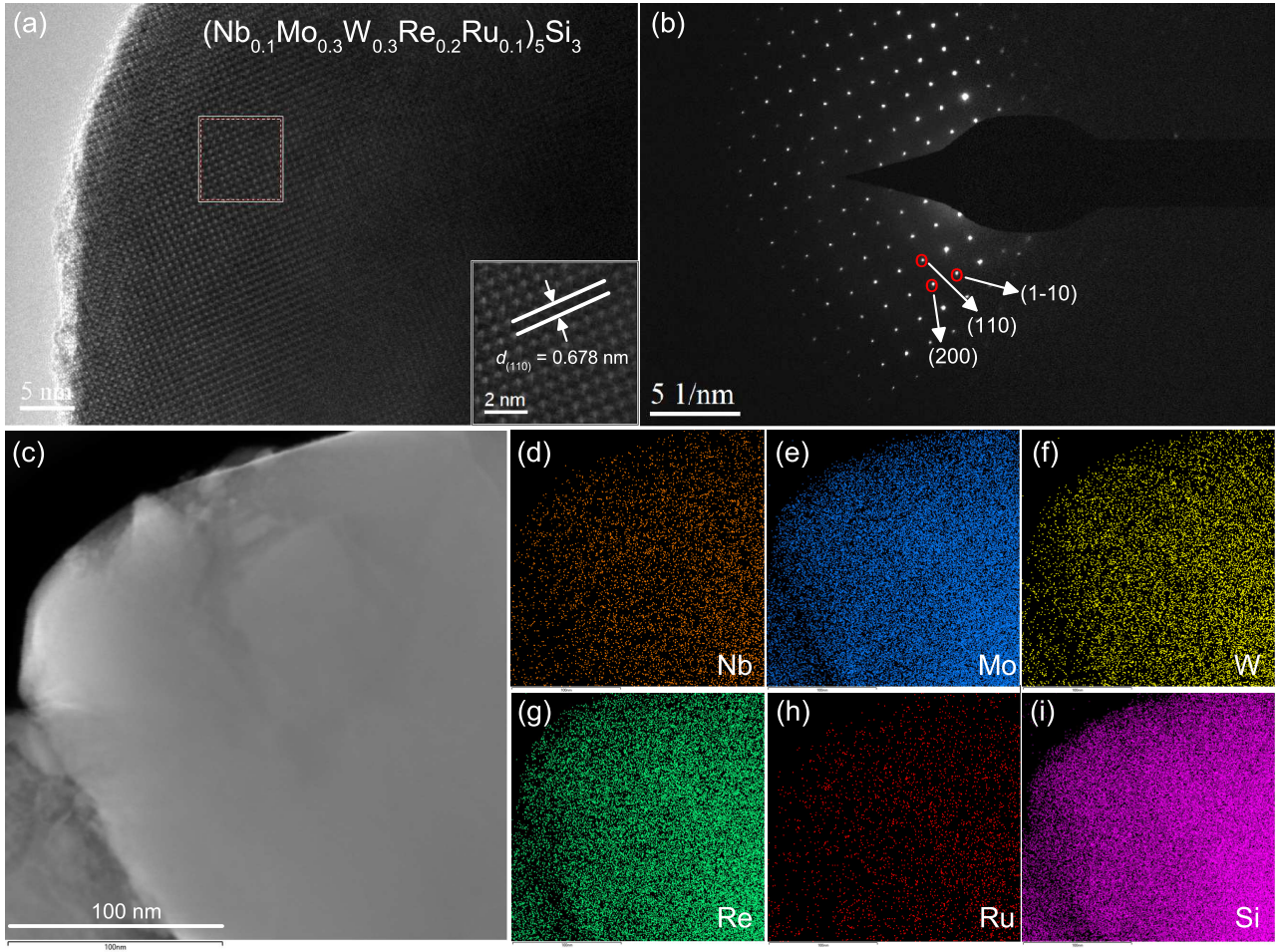}
	\caption{
	(a, b) High resolution TEM image and corresponding SAED pattern, respectively, for (Nb$_{0.1}$Mo$_{0.3}$W$_{0.3}$Re$_{0.2}$Ru$_{0.1}$)$_{5}$Si$_{3}$.
    The inset of panel (a) shows a zoom of the image and the lattice spacing is labeled.
    (c) TEM image for this HES with a scale bar of 100 nm. The corresponding elemental maps of Nb, Mo, W, Re, Ru and Si are shown in (d-i).
	}
	\label{fig3}
\end{figure*}

In the unit cell of W$_{5}$Si$_{3}$-type $M_{5}$Si$_{3}$, there are two distinct cation $M$ sites: $M$1(0.074, 0.223, 0) and $M$2(0, 0.5, 0.25).
Therefore it is prudent to examine the cation distribution in (Nb$_{0.1}$Mo$_{0.3}$W$_{0.3}$Re$_{0.2}$Ru$_{0.1}$)$_{5}$Si$_{3}$ and (Nb$_{0.2}$Mo$_{0.3}$W$_{0.3}$Re$_{0.1}$Ru$_{0.1}$)$_{5}$Si$_{3}$, in particular considering that cation ordering has been observed in Mn$_{5}$Si$_{3}$-type (V$_{1/5}$Cr$_{1/5}$Nb$_{1/5}$Ta$_{1/5}$W$_{1/5}$)$_{5}$Si$_{3}$ \cite{ref24}.
To this end, we compare the experimental XRD pattern of (Nb$_{0.2}$Mo$_{0.3}$W$_{0.3}$Re$_{0.1}$Ru$_{0.1}$)$_{5}$Si$_{3}$ with the calculated ones assuming different cation distributions in Fig. 1(b).
In the disordered model, the Nb, Mo, W, Re and Ru atoms are assumed to occupy randomly both cation sites.
In the ordered model, the cations in the first and second parentheses are assumed to occupy the $M$1 and $M$2 sites, respectively.
Actually, we have calculated the XRD patterns for thirteen different cation ordered distributions. However, due to the close atomic numbers of Nb and Mo as well as W and Re, it is difficult to distinguish between these pairs of elements by x-ray diffraction. Hence the patterns are more or less similar and only four representative ones are shown here.
One can see that the cation disordered model provides the best agreement with the experimental observation, and the apparent difference in the (002) peak intensity is presumably due to the preferential orientation.
In contrast, the calculated intensities of the (110) and (200) reflections are either too strong for (Re$_{0.1}$Ru$_{0.1}$)$_{5}$(Nb$_{0.2}$Mo$_{0.3}$W$_{0.3}$)$_{5}$Si$_{3}$ and (W$_{0.2}$)$_{5}$(Nb$_{0.2}$Mo$_{0.3}$W$_{0.1}$Re$_{0.1}$Ru$_{0.1}$)$_{5}$Si$_{3}$ or two weak for (Mo$_{0.2}$)$_{5}$(Nb$_{0.2}$Mo$_{0.1}$W$_{0.3}$Re$_{0.1}$Ru$_{0.1}$)$_{5}$Si$_{3}$ and (Nb$_{0.2}$)$_{5}$(Mo$_{0.3}$W$_{0.1}$Re$_{0.1}$Ru$_{0.1}$)$_{5}$Si$_{3}$.
The disorder model is further corroborated by the structural refinement result shown in \cite{SM} using the JANA2006 programme.
Indeed, all the diffraction peaks can be well fitted with small reliability factors $R_{\rm wp}$ = 8.4\% and $R_{\rm p}$ = 5.9\%.
Hence, the classification of both (Nb$_{0.1}$Mo$_{0.3}$W$_{0.3}$Re$_{0.2}$Ru$_{0.1}$)$_{5}$Si$_{3}$ and (Nb$_{0.2}$Mo$_{0.3}$W$_{0.3}$Re$_{0.1}$Ru$_{0.1}$)$_{5}$Si$_{3}$ as HESs is reasonable,
although the possibility of short-range cation ordering cannot be excluded completely.

The morphology and chemical composition of both (Nb$_{0.1}$Mo$_{0.3}$W$_{0.3}$Re$_{0.2}$Ru$_{0.1}$)$_{5}$Si$_{3}$ and (Nb$_{0.2}$Mo$_{0.3}$W$_{0.3}$Re$_{0.1}$Ru$_{0.1}$)$_{5}$Si$_{3}$ are examined by SEM and EDX measurements, whose results are displayed in Fig. 2. From Figure 2(a), one can see that the (Nb$_{0.1}$Mo$_{0.3}$W$_{0.3}$Re$_{0.2}$Ru$_{0.1}$)$_{5}$Si$_{3}$ HES is free from pores and voids on the scale of 2.5 $\mu$m. While a few small particles with size below 1 $\mu$m are also present,
the elemental maps shown in Figs. 2(b-g) reveal a uniform distribution of the Nb, Mo, W, Re, Ru, and Si elements. The situation is very similar for (Nb$_{0.2}$Mo$_{0.3}$W$_{0.3}$Re$_{0.1}$Ru$_{0.1}$)$_{5}$Si$_{3}$, as can be see in Figs. 2(h-n).
By the EDX analysis, the actual chemical compositions are determined to be (Nb$_{0.13}$Mo$_{0.28}$W$_{0.29}$Re$_{0.19}$Ru$_{0.11}$)$_{4.98}$Si$_{3.02}$ and (Nb$_{0.20}$Mo$_{0.27}$W$_{0.29}$Re$_{0.10}$Ru$_{0.14}$)$_{5.02}$Si$_{2.98}$ for the former and the latter, respectively, indicating that both HESs are stoichiometric within the measurement uncertainty of 3\%.

\begin{figure*}
	\includegraphics*[width=17.6cm]{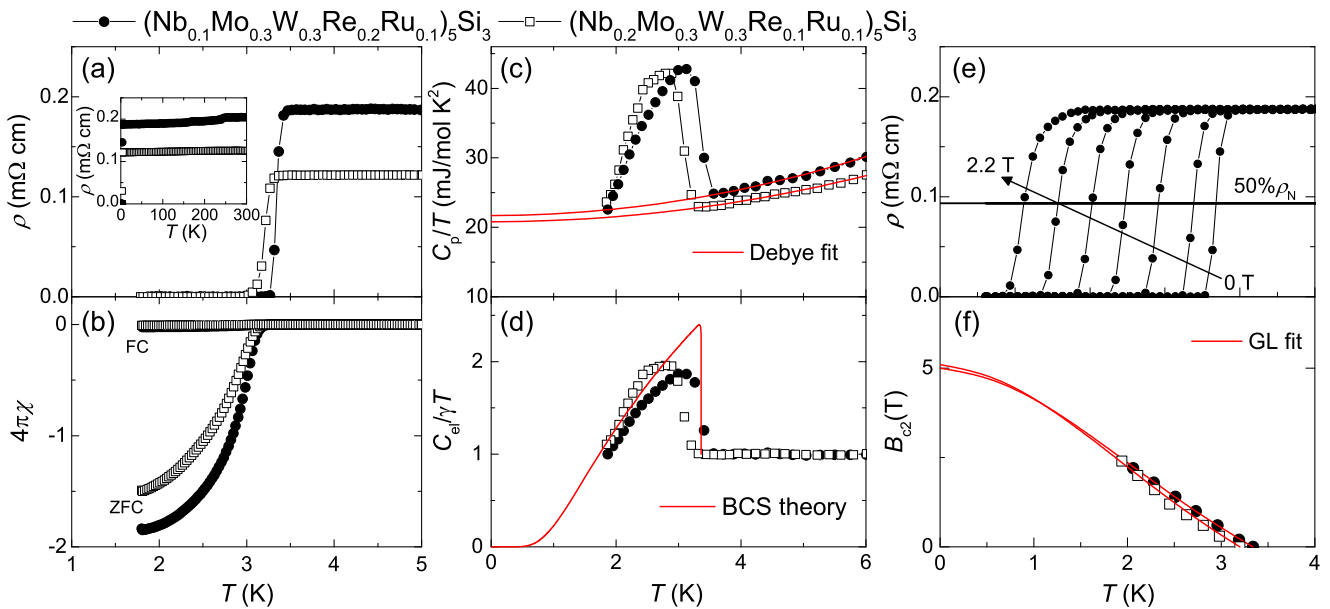}
	\caption{
		(a, b) Low temperature resistivity and magnetic susceptibility for (Nb$_{0.1}$Mo$_{0.3}$W$_{0.3}$Re$_{0.2}$Ru$_{0.1}$)$_{5}$Si$_{3}$ and (Nb$_{0.2}$Mo$_{0.3}$W$_{0.3}$Re$_{0.1}$Ru$_{0.1}$)$_{5}$Si$_{3}$. The inset in panel (a) shows the resistivity data up to 300 K. (c) Temperature dependence of $C_{\rm p}$/$T$ for the two HESs. The solid lines are Debye fits to the normal-state data. (d) Normalized electronic specific heat plotted as a function of temperature for the two HESs. The solid curve denotes the expected behavior from the BCS theory. (e) Resistive transition under increasing fields for (Nb$_{0.1}$Mo$_{0.3}$W$_{0.3}$Re$_{0.2}$Ru$_{0.1}$)$_{5}$Si$_{3}$. The horizontal line is a guide to the eyes and the arrow marks the direction of field increasing. (f) Upper critical field versus temperature phase diagrams for the two HESs.
The solid lines are fits to the data by the GL model.
	}
	\label{fig4}
\end{figure*}
Figure 3(a) shows the high-resolution TEM (HRTEM) image with a scale bar of 5 nm for (Nb$_{0.1}$Mo$_{0.3}$W$_{0.3}$Re$_{0.2}$Ru$_{0.1}$)$_{5}$Si$_{3}$ taken along the [0 0 1] zone axis.
The image shows clear lattice fringes, indicating a highly crystalline nature.
On zooming-in [the inset of Fig. 3(a)], one can see that the nearest neighbour atoms form square lattices with a spacing of 0.678 nm,
which corresponds to (110) and (1-10) planes of the W$_{5}$Si$_{3}$-type lattice.
The corresponding selected-area electron diffraction (SAED) exhibits a well-defined spot pattern, and the spots marked by the arrows can be indexed to the (110), (1-10) and (200) planes.
The TEM image and EDX elemental maps with a scale bar of 100 nm for this HES is displayed in Fig. 3(c-i).
It is obvious that all the constitute elements are uniformly distributed within the experimental resolution.
These results, together with the above ones, confirm that the HES is homogeneous on both a macroscopic and microscopic level.

We then turn the attention to the physical properties of (Nb$_{0.1}$Mo$_{0.3}$W$_{0.3}$Re$_{0.2}$Ru$_{0.1}$)$_{5}$Si$_{3}$ and (Nb$_{0.2}$Mo$_{0.3}$W$_{0.3}$Re$_{0.1}$Ru$_{0.1}$)$_{5}$Si$_{3}$, whose resistivity and magnetic susceptibility are displayed in Figs. 4(a) and (b), respectively. For both HESs, the resistivity exhibits a nearly temperature independent behaviour [the inset of Fig. 4(a)] and drops rapidly to zero at low temperatures, which is accompanied by a clear diamagnetic transition.
At 1.8 K, the zero-field cooling susceptibility data correspond to large shield fractions of $\sim$150-180\% without correction for the demagnetization factor.
These results provide clear evidence for a bulk superconducting transition in the HESs.
Here $T_{\rm c}$ is defined as the temperature for 50\%$\rho_{\rm N}$, where $\rho_{\rm N}$ is the normal-state resistivity just above the transition.
This gives $T_{\rm c}$ = 3.3 K and 3.2 K for (Nb$_{0.1}$Mo$_{0.3}$W$_{0.3}$Re$_{0.2}$Ru$_{0.1}$)$_{5}$Si$_{3}$ and (Nb$_{0.2}$Mo$_{0.3}$W$_{0.3}$Re$_{0.1}$Ru$_{0.1}$)$_{5}$Si$_{3}$, respectively.
Intriguingly, these $T_{\rm c}$ values are higher than individual binary counterparts $\beta$-Nb$_{5}$Si$_{3}$ ($T_{\rm c}$ = 0.7 K) and W$_{5}$Si$_{3}$ ($T_{\rm c}$ = 2.7 K).

\begin{figure*}
	\includegraphics*[width=17.6cm]{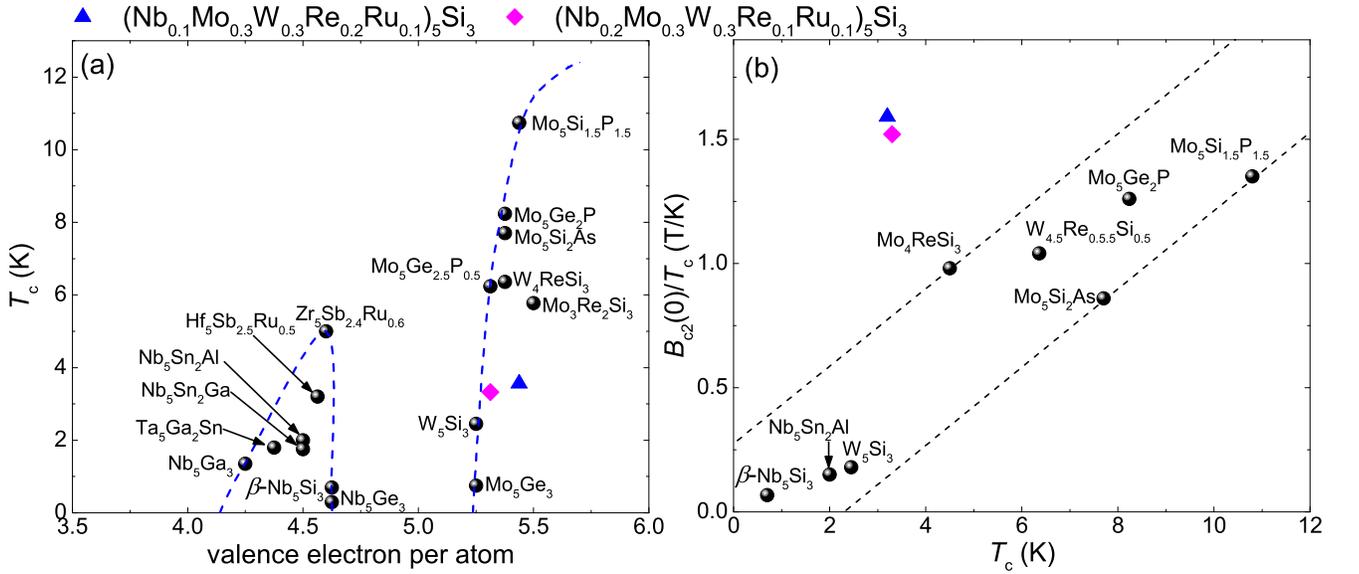}
	\caption{
	 (a) Dependence of $T_{\rm c}$ on the average number of valence electrons per atom ratio for various W$_{5}$Si$_{3}$-type superconductors. (b) $B_{\rm c2}$(0)/$T_{\rm c}$ of typical superconductors of this family plotted as a function of $T_{\rm c}$.
In both panels, the data points for (Nb$_{0.1}$Mo$_{0.3}$W$_{0.3}$Re$_{0.2}$Ru$_{0.1}$)$_{5}$Si$_{3}$ and (Nb$_{0.2}$Mo$_{0.3}$W$_{0.3}$Re$_{0.1}$Ru$_{0.1}$)$_{5}$Si$_{3}$ are shown in different symbols and colors, and the dashed lines are guides to the eyes.
	}
	\label{fig5}
\end{figure*}
Figure 4(c) shows the specific heat ($C_{\rm p}$) results plotted as $C_{\rm p}$/$T$ versus $T$.
A clear jump is observed for both HESs, confirming the bulk nature of superconductivity.
Analyzing the normal-state data using the Debye model,
\begin{equation}
C_{\rm p}/T = \gamma + \beta T^{2},
\end{equation}
yields the Sommerfeld coefficient $\gamma$ = 21.7 and 20.8 mJ/molK$^{2}$ and the phonon specific heat coefficient $\beta$ = 0.2353 and 0.1842 mJ/molK$^{4}$ for (Nb$_{0.1}$Mo$_{0.3}$W$_{0.3}$Re$_{0.2}$Ru$_{0.1}$)$_{5}$Si$_{3}$ and (Nb$_{0.2}$Mo$_{0.3}$W$_{0.3}$Re$_{0.1}$Ru$_{0.1}$)$_{5}$Si$_{3}$, respectively.
The $\beta$ values allow us to calculate the Debye temperature $\Theta_{\rm D}$ = 404 K and 472 K for the former and latter HESs, respectively, according to the equation,
\begin{equation}
\Theta_{\rm D} = (12\pi^{4}NR/5\beta)^{1/3},
\end{equation}
where $N$ = 8 and $R$ = 8.314 J/molK is the molar gas constant.
Then the electronic specific heat is obtained as $C_{\rm el}$ = $C_{\rm p}$ $-$ $\beta$$T^{3}$ and plotted as $C_{\rm el}$/$\gamma$$T$ versus $T$ in Fig. 4(d).
From entropy conserving construction \cite{entropy}, the $T_{\rm c}$ is found to be 3.3 K and 3.0 K for the (Nb$_{0.1}$Mo$_{0.3}$W$_{0.3}$Re$_{0.2}$Ru$_{0.1}$)$_{5}$Si$_{3}$ and (Nb$_{0.2}$Mo$_{0.3}$W$_{0.3}$Re$_{0.1}$Ru$_{0.1}$)$_{5}$Si$_{3}$ HES, respectively, which are the same as or slightly lower than those determined by resistivity measurements. Furthermore, their specific heat jumps are estimated to be 1.1-1.2, considerably smaller than the BCS value of 1.43 \cite{ref28}.
Indeed, the temperature dependence of $C_{\rm el}$/$\gamma$$T$ exhibits deviation from the BCS theory \cite{ref28}. This behavior is reminiscent of those observed in $\alpha$-Mn-type Nb$_{25}$Mo$_{5+x}$Re$_{35}$Ru$_{25-x}$Rh$_{10}$ \cite{alphaMn} and bcc-type (NbTa)$_{0.67}$(MoHfW)$_{0.33}$ \cite{bccHEA} HEA superconductors, which is likely due to either the presence of multiple gaps or the broadening of the transition at $T_{\rm c}$ caused by the strong atomic disorder inherent to the high-entropy systems.
On the other hand, the electron-phonon coupling strength $\lambda_{\rm ep}$ is found to 0.51$-$0.53 based on the inverted McMillan formula \cite{ref30},
\begin{equation}
\lambda_{\rm ep}=\frac{1.04+\mu^{*} \ln \left(\Theta_{\rm D} / 1.45 T_{\mathrm{c}}\right)}{\left(1-0.62 \mu^{*}\right) \ln \left(\Theta_{\rm D} / 1.45 T_{\mathrm{c}}\right)-1.04},
\end{equation}
where $\mu^{\ast}$ = 0.13 is the Coulomb repulsion pseudopotential.
This is consistent with the reduced specific heat jump and unveils a weakly coupled superconducting state in these HESs.

The upper critical field $B_{\rm c2}$ is determined by resistivity measurements under magnetic fields and the results for (Nb$_{0.1}$Mo$_{0.3}$W$_{0.3}$Re$_{0.2}$Ru$_{0.1}$)$_{5}$Si$_{3}$ are shown in Fig. 4(e).
With increasing field from 0 T to 2.2 T, the resistive transition is suppressed toward lower temperatures, as expected for a superconducting transition.
The $T_{\rm c}$ for each field is defined by the same 50\%$\rho_{\rm N}$ criterion as above and the constructed $B_{\rm c2}$ versus $T$ phase diagrams are displayed in Fig. 4(f).
Similar to Mo$_{5-x}$Re$_{x}$Si$_{3}$ \cite{ref23} and W$_{5-x}$Re$_{x}$Si$_{3}$ \cite{ref27}, the $B_{\rm c2}$($T$) data well obey the Ginzburg-Landau model \cite{ref31}
\begin{equation}
B_{\rm c2}(T) = B_{\rm c2}(0) \frac{1-t^{2}}{1+t^{2}},
\end{equation}
where $B_{\rm c2}$(0) is the zero-temperature upper critical field and $t$ = $T$/$T_{\rm c}$ is the reduced temperature.
Extrapolation to 0 K yields $B_{\rm c2}$(0) = 5.0 T and 5.1 T for (Nb$_{0.1}$Mo$_{0.3}$W$_{0.3}$Re$_{0.2}$Ru$_{0.1}$)$_{5}$Si$_{3}$ and (Nb$_{0.2}$Mo$_{0.3}$W$_{0.3}$Re$_{0.1}$Ru$_{0.1}$)$_{5}$Si$_{3}$, respectively.
Once $B_{\rm c2}$(0) is known, the Ginzburg-Landau (GL) coherence length $\xi_{\rm GL}$(0) is calculated to be 8.1 nm and 8.0 nm for the former and the latter, respectively, based on the equation
\begin{equation}
\xi_{\rm GL}(0) = \sqrt{\frac{\Phi_{0}}{2\pi B_{\rm c2}(0)}},
\end{equation}
where $\Phi_{0}$ = 2.07 $\times$ 10$^{-15}$ Wb is the flux quantum.

For binary and ternary W$_{5}$Si$_{3}$-type superconductors,
there exits a correlation between their $T_{\rm c}$ and the average number of valence electrons per atom ratio ($e$/$a$) \cite{ref23}.
As illustrated in Fig. 5(a), two superconducting regions are observed in the $e$/$a$ ranges of 4.25-4.625 and 5.24-5.43.
Specifically, $T_{\rm c}$ exhibits a maximum at $e$/$a$ = 4.6 in the former range while increases steeply with $e$/$a$ in the latter one.
The $e$/$a$ values of the (Nb$_{0.1}$Mo$_{0.3}$W$_{0.3}$Re$_{0.2}$Ru$_{0.1}$)$_{5}$Si$_{3}$ ($e$/$a$ = 5.43) and (Nb$_{0.2}$Mo$_{0.3}$W$_{0.3}$Re$_{0.1}$Ru$_{0.1}$)$_{5}$Si$_{3}$ ($e$/$a$ = 5.31) HESs fall within the latter range, which provides a natural explanation about the observed superconductivity.
Nevertheless, while the $T_{\rm c}$ dependence on $e$/$a$ for (Nb$_{0.2}$Mo$_{0.3}$W$_{0.3}$Re$_{0.1}$Ru$_{0.1}$)$_{5}$Si$_{3}$ follows well the previous trend, $T_{\rm c}$ of (Nb$_{0.1}$Mo$_{0.3}$W$_{0.3}$Re$_{0.2}$Ru$_{0.1}$)$_{5}$Si$_{3}$ is only one-third that of Mo$_{5}$Si$_{1.5}$P$_{1.5}$ despite their almost identical $e$/$a$ \cite{ref32}.
Hence it appears that, other than $e$/$a$, the elemental composition also plays a role in determining the $T_{\rm c}$ \cite{note1}.

Finally, we present a comparison between the $B_{\rm c2}$(0)/$T_{\rm c}$ ratios of typical W$_{5}$Si$_{3}$-type superconductors, which are plotted as a function of $T_{\rm c}$ in Fig. 5(b).
It is noteworthy that the $B_{\rm c2}$(0)/$T_{\rm c}$ of binary and ternary superconductors tends to increase with increasing $T_{\rm c}$.
For $\beta$-Nb$_{5}$Si$_{3}$ with the lowest $T_{\rm c}$ of 0.7 K, its $B_{\rm c2}$(0)/$T_{\rm c}$ is only 0.067 T/K.
In comparison, Mo$_{5}$Si$_{1.5}$P$_{1.5}$ with the highest $T_{\rm c}$ of 10.8 K has a $B_{\rm c2}$(0)/$T_{\rm c}$ ratio of 1.35, which is more than twenty times larger than that of $\beta$-Nb$_{5}$Si$_{3}$.
Remarkably, despite their low $T_{\rm c}$, the $B_{\rm c2}$(0)/$T_{\rm c}$ ratios of the (Nb$_{0.1}$Mo$_{0.3}$W$_{0.3}$Re$_{0.2}$Ru$_{0.1}$)$_{5}$Si$_{3}$ and (Nb$_{0.2}$Mo$_{0.3}$W$_{0.3}$Re$_{0.1}$Ru$_{0.1}$)$_{5}$Si$_{3}$ HESs reach above 1.5.
These values are not only more than 1.5 times than that (= 0.98) of Mo$_{4}$ReSi$_{3}$ with a similar $T_{\rm c}$ but also the highest among known W$_{5}$Si$_{3}$-type superconductors.
In the dirty limit, $B_{\rm c2}$(0)/$T_{\rm c}$ is shown to be proportional to $\gamma$$\rho_{\rm N}$ \cite{ref33}.
Since the $\gamma$ values of the HESs are close to that of Mo$_{4}$ReSi$_{3}$, it is reasonable to speculate that the increases in $\rho_{\rm N}$ due to the increased disorder is mainly responsible for the enhancement of $B_{\rm c2}$(0)/$T_{\rm c}$ in the HES case.
Nevertheless, it is difficult to quantify the intrinsic $\rho_{\rm N}$ values since the HESs are of a polycrystalline nature and their resistivity is mainly determined by the contribution from grain boundaries.
Note that this enhancement is also observed in the (V$_{0.5}$Nb$_{0.5}$)$_{3-x}$Mo$_{x}$Al$_{0.5}$Ga$_{0.5}$ \cite{ref34} and interstitial Cr$_{5+x}$Mo$_{35-x}$W$_{12}$Re$_{35}$Ru$_{13}$C$_{20}$ HEAs \cite{ref35,note2}, which calls for further studies to assess its generality in other high-entropy superconductors.

\section{Conclusion}
In summary, we synthesized and characterized two new (Nb$_{0.1}$Mo$_{0.3}$W$_{0.3}$Re$_{0.2}$Ru$_{0.1}$)$_{5}$Si$_{3}$ and (Nb$_{0.2}$Mo$_{0.3}$W$_{0.3}$Re$_{0.1}$Ru$_{0.1}$)$_{5}$Si$_{3}$ HESs.
It is found that both HESs possess the W$_{5}$Si$_{3}$-type structure with a disordered cation distribution and become bulk superconductors on cooling below 3.2-3.3 K, which are higher than those of $\beta$-Nb$_{5}$Si$_{3}$ and W$_{5}$Si$_{3}$.
Moreover, the analysis of their specific heat jumps point to a weakly coupled superconducting state.
While these HESs have a similar $e/a$ dependence of $T_{\rm c}$ to that of the isostructural binary and ternary counterparts, their $B_{\rm c2}$(0)/$T_{\rm c}$ ratios are the largest among this family of superconductors.
Our results not only expand the structure diversity of HECs but also present the first superconducting high-entropy nonoxide ceramics, which may help to advance the understanding and design of these multicomponent materials.

\section*{ACKNOWLEDGEMENT}
We acknowledge financial support by the foundation of Westlake University and the Service Center for Physical Sciences for technical assistance in SEM measurements.
The works at Zhejiang University and Kunming University of Science and Technology are supported by the National Natural Science Foundation of China (12050003) and the Yunnan Fundamental Research Projects (202201AU070118), respectively.


\begin{thebibliography}{31}
\bibitem{note1}No systematic trend is found bewteen the $T_{\rm c}$ and unit-cell volume for W$_{5}$Si$_{3}$-type superconductors.
\bibitem{note2}Since carbon occupies the interstital sites, the Cr$_{5+x}$Mo$_{35-x}$W$_{12}$Re$_{35}$Ru$_{13}$C$_{20}$ series are classifed as interstital HEAs rather than high-entropy carbides.
\end{thebibliography}
\end{document}